\documentclass[notitlepage,superscriptaddress,nofootinbib,amsmath,longbibliography,amssymb,11pt,groupedaddress]{revtex4-1}
\usepackage{amsmath,amsthm}
\usepackage{graphicx}
\usepackage[colorlinks=true, linkcolor=blue, citecolor=magenta, hyperfootnotes=true]{hyperref}
\hypersetup{
	colorlinks   = true, 
	urlcolor     = blue, 
	linkcolor    = blue, 
	citecolor    = red   
}
\usepackage[normalem]{ulem}
\usepackage{multirow}
\makeatletter
\gdef\@fpheader{ }
\makeatother

\newcommand{\nc}{\newcommand}
\nc{\ba}{\begin{eqnarray}}
\nc{\ea}{\end{eqnarray}}
\newcommand{\calR}{{\cal{R}}}

\newcommand{\calP}{{\cal{P}}}

\nc{\bfx}{{\bf{x}}}
\nc{\bfk}{{\bf{k}}}
\nc{\Mpl}{{M_{\text{pl}}}}
\nc{\Mplsq}{{M^2_{\text{pl}}}}
\nc{\fNL}{{f_{\text{NL}}}}

\nc{\mhn}{\color{blue}{\bf MHN: }}
\nc{\mth}{\theta_{\text{th}}}

\usepackage[toc,page]{appendix}

\begin{document}

\title{ One consistency relation for all single-field inflationary models}

\author{Mohammad Hossein Namjoo}
\email{mh.namjoo@ipm.ir}
\affiliation{School of Astronomy, Institute for Research in Fundamental Sciences (IPM),\\ Tehran, Iran, P.O. Box 19395-5531}


\begin{abstract}
In this paper, we present a non-Gaussianity consistency relation that enables the calculation of the squeezed limit bispectrum of the curvature perturbation in single-field inflationary models by carefully inspecting the background evolution and the linear perturbation theory. The consistency relation is more general than others in the literature since it does not require any specific symmetry, conservation of the curvature perturbation at large scales, attractor background evolution or canonical kinetic energy of the inflaton field. We demonstrate that  all known examples of the squeezed limit bispectrum in single-field models of inflation can be  reproduced within this framework. 
\end{abstract}

\maketitle
\tableofcontents

\newpage 

\section{Introduction}
Consistency relations are generic expressions that relate  higher order correlation functions to  lower order ones. In cosmology, they are of significant interest for both theoretical and observational reasons. On the theoretical side, consistency relations indicate the presence of specific symmetries \cite{Assassi:2012zq,Hui:2018cag,Hinterbichler:2012nm,Hinterbichler:2013dpa} or specific conditions under which the universe evolved \cite{Maldacena:2002vr,Creminelli:2013nua,Pajer:2017hmb}. They also allow one to compute higher order correlation functions by simpler means, as the direct calculation can sometimes be tedious. On the observational side, any observed violation of each consistency relation implies the violation of its underlying assumptions. This gives us valuable information about the evolution of the universe. 

The consistency conditions that are based on symmetries can be viewed as signatures of these symmetries and are thus of significant value. However, for exactly the same reason, they are restrictive and if the assumed symmetries are broken, these consistency relations fail to make correct predictions. In this paper, we present a consistency relation that avoids such assumptions (as well as other assumptions about the background evolution or the form of the Lagrangian) and is, therefore, more generic. As we will see, it allows us to compute the squeezed limit three point correlation functions in intrinsically different single-field scenarios without the need for knowledge beyond the linear perturbation theory. 

We present our results in terms of the curvature perturbation on comoving slices which we denote by $\calR$. 
The power spectrum of the curvature perturbation is expressed by $\langle \calR_{\bfk_1}\calR_{\bfk_2} \rangle =(2\pi)^3 \delta^3(\bfk_1+\bfk_2)  P_{\calR_{k_1}} $. We occasionally use the dimensionless form of the power spectrum which is defined by $\calP_{\calR_k} =\frac{k^3}{2\pi^2} P_{\calR_k}$.
The bispectrum of the curvature perturbation is given by
\begin{equation}
\label{eq: bispectrum}
\left\langle\mathcal{R}_{\mathbf{k}_{\mathbf{1}}} \mathcal{R}_{\mathbf{k}_{\mathbf{2}}} \mathcal{R}_{\mathbf{k}_{\mathbf{3}}}\right\rangle \equiv(2 \pi)^3 \delta^{(3)}\left(\mathbf{k}_{\mathbf{1}}+\mathbf{k}_{\mathbf{2}}+\mathbf{k}_{\mathbf{3}}\right) B_{\mathcal{R}}\left(k_1, k_2, k_3\right) .
\end{equation}
We are interested in the squeezed limit of the bispectrum where one mode has a much longer wavelength than the other two and which we quantify by
\begin{equation}
\label{eq: fnl}
B_{\mathcal{R}}\left(k_s, k_s, k_\ell \right) \to (2 \pi)^4 \frac{1}{k_s^3 k_\ell^3} \calP_{\mathcal{R}_{s}} \calP_{\mathcal{R}_{\ell}}\, \frac{3}{5} f_{\mathrm{NL}}\, , \qquad \text{for $k_\ell/k_s\to 0$}\, ,
\end{equation}
where $k_\ell$ and $k_s$ are the wavenumbers of the long mode and the short modes, respectively, and --- to simplify our notation --- we use $\calR_\ell$ and $\calR_s$ as shorthands for  $\calR_{k_\ell}$ and $\calR_{k_s}$. (When an expression holds for both long and short modes, we denote the corresponding wavenumber as $k$.)
Eq.~\eqref{eq: fnl} also defines $\fNL$ which measures the size of non-Gaussianity in the squeezed limit.

 The direct way to calculate the bispectrum involves  computing the Lagrangian for the fluctuations up to the third order. The purpose of this paper is to derive  expressions for the squeezed limit bispectrum by solely  analyzing  the background evolution and the linear perturbation theory. 
\section{The consistency relation in its most general form}

In this section, we present the consistency relation in its most general and abstract form which will be used in different situations in Sec.~\ref{examples}.
We start by noting that for single-field models when there is a hierarchy in scales $k_\ell \ll k_s$ we may write \cite{Creminelli:2004yq}
\ba 
\label{long_short_3pt}
\langle \calR_\ell(\tau) \calR_s(\tau)  \calR_s(\tau) \rangle \simeq \langle \calR_\ell(\tau)\, \langle \calR_s(\tau)  \calR_s(\tau)\rangle_{\calR_\ell} \rangle 
\ea 
where $\tau$ is the conformal time and $\langle \calR_s^2 \rangle_{\calR_\ell}$ is the two-point correlation function of the short modes under the influence of the long mode. The intuition behind  Eq.~\eqref{long_short_3pt}, in accord with the separate universe picture, is that the short modes perceive the long mode as a change in their background evolution. The important assumption behind the validity of this expression is that $\calR_\ell$ can be treated classically (which is justified in a quasi-de Sitter background and as long as $\calR_\ell$ is a super-horizon mode). However,  notice that the conservation of $\calR_\ell$ (which is violated for non-attractor evolutions) is not required. Since the interaction is small in an inflationary background, this expression can be Taylor expanded. A crucial point to notice is that since we allow  $\calR_\ell$ to vary in time, its effect must be taken into account over the entire history of the short modes. Thus, we make a {\it functional Taylor expansion} as follows

\ba 
\langle \calR_s^2(\tau) \rangle_{\calR_\ell} \simeq \langle \calR_s^2 \rangle_{0} + \int_{\tau_\ell}^{\tau} \calR_\ell(\tau') \dfrac{\delta }{\delta \calR_\ell(\tau') } \langle \calR_s^2(\tau) \rangle \, d\tau' \, ,
\ea 
where we have neglected  higher order terms in the expansion and $\tau_\ell$ is an early-time, after the horizon-crossing of $\calR_\ell$.  We are mainly interested in the correlations at late-times; so we will take the limit $\tau \to 0$. This relation assumes that all correlations between the long mode and the short modes occur after $\tau_\ell$. This is justified as long as the sub-horizon modes are in the Bunch-Davies vacuum.  Substituting this relation to Eq.~\eqref{long_short_3pt} results in 
\ba 
\label{main}
B(k_s,k_s,k_\ell) \simeq  \int_{\tau_\ell}^{\tau} \langle \calR_\ell(\tau) \calR_\ell(\tilde \tau) \rangle' \dfrac{\delta }{\delta \calR_\ell(\tilde \tau) } \langle \calR_s^2(\tau) \rangle'  \, d\tilde \tau \, ,
\ea 
where $\langle . \rangle'$ denotes the same correlation function $\langle . \rangle$ with the factors of $(2\pi)^3$ and the momentum conserving delta-function removed.  
This is our key relation that, we contend, reproduces all known  squeezed limit bispectra in single-field models of inflation. To utilize this relation, one needs to inspect how the late-time two-point correlation function of the short modes depends on the history of the evolution and then study how this history is affected by the long mode.  Care must be taken to when each variable is computed especially if $\calR_\ell$ is not conserved. We will see that the late-time power spectrum is only sensitive to a few instances of time in its past so that Eq.~\eqref{main}  greatly simplifies in specific scenarios. It is worth noting that this relation does not assume any symmetry, nor does it assume the conservation of $\calR_\ell$. Therefore, one can expect that this finding will result in the true bispectrum in all single-field models of inflation.

Let us make explicit the assumptions behind our key relation Eq.~\eqref{main}. We require that (i) there is a hierarchy between scales ($k_\ell \ll k_s$); (ii) the long mode  behaves classically; and (iii) the effect of $\calR_\ell$ up to the linear order suffices. Note that (iii) is used both  explicitly and implicitly in Eq.~\eqref{main}. We have neglected higher order terms in the Taylor expansion. Moreover, we did not take functional derivatives with respect to $\dot \calR_\ell$ which, generally, is an independent degree of freedom. However, since we are working up to the linear order in $\calR_\ell$ and under condition (ii), one can remove $\dot \calR_\ell$ in favor of $\calR_\ell$. This is possible as a consequence of the decay of one independent solution at large scales relative to the other, which also leads $\calR_\ell$ to become classical (requirement (ii)).  If in some situation the condition (iii) is not satisfied, it is straightforward to extend Eq.~\eqref{main} to also include the functional derivatives with respect to $\dot \calR_\ell$ as well as higher order terms in the Taylor expansion. It is sufficient for this paper to consider only the restrictive form of Eq.~\eqref{main} and neglect other contributions.\footnote{We thank Enrico Pajer for a discussion on this issue.} 

As a final remark, it is worth noting that Eq.~\eqref{main} differs from many other consistency relations in the following sense. While other consistency relations usually connect observables,  employing Eq.~\eqref{main} requires some knowledge of the history of the evolution. Nonetheless, Eq.~\eqref{main} may still be viewed as a consistency condition since, as we will see shortly, it allows one to construct the squeezed limit bispectrum by analyzing the linear fluctuations.\footnote{We thank Mehrdad Mirbabayi and Enrico Pajer for their correspondence on this matter.}

\section{Application of the consistency relation to specific scenarios}
\label{examples}

In this section, we demonstrate the generality of  the key relation Eq.~\eqref{main} by reproducing several known squeezed limit bispectra from intrinsically different single-field models. 

\subsection{Slow-roll inflation}
\label{SR}
We start with the simplest example, i.e., the slow-roll inflationary model. We consider a general Lagrangian of the form $L(X,\phi)$ where $X=\frac12 \partial_\mu \phi \partial^\mu \phi$; the power spectrum is  given by \cite{Chen:2006nt}
\ba 
\label{PR_SR}
\calP_{\calR_s}(0) =   \dfrac{H_*^2}{8\pi^2 \Mpl^2 \epsilon_*  c_{s_*}} \, 
\ea 
where $\epsilon\equiv -\dot H/H^2=XL_{,X}/\Mplsq H^2$, $c_s \equiv  L_{,X}/(L_{,X}+2X L_{,XX})$, $\Mpl$ is the reduced Planck mass and the subscript $*$ denotes that the quantities must be computed at the horizon-crossing time $k_s=a_* H_*$. In this expression, we have made explicit that it holds at late times, i.e., $\tau \to 0$ even though different quantities are computed at the horizon-crossing time.  This is because in slow-roll inflation, the curvature perturbation is conserved after horizon-crossing.  To obtain the consistency relation, we ask how Eq.~\eqref{PR_SR} depends on the history of the evolution. Notice that $H$, $\epsilon$ and $c_s$ are generally functions of time. Thus, the history comes in via (i) the initial conditions and (ii) the horizon-crossing time. (i) is negligible here since we are dealing with the slow-roll  models of inflation where the attractor behavior of the background makes the final results insensitive to the initial conditions. (ii) can be taken into account by noting that the line element for the short modes in the presence of the long mode is $ds^2\simeq -dt^2 + a^2 e^{2\calR_\ell} d\bfx^2$. Thus, the effect of the long mode is to rescale the scale factor and we have 
\ba 
\label{func_deriv_SR}
\calR_\ell(\tilde \tau)\dfrac{\delta }{\delta \calR_\ell(\tilde \tau) } \langle \calR_s^2(0) \rangle'  \simeq - \calR_\ell(0)\dfrac{dP_{\calR_s}}{d\ln a_*}\delta(\tilde \tau-\tau_*)  \simeq   - \calR_\ell(0) P_{\calR_s} (n_s-1) \delta(\tilde \tau-\tau_*) \, ,
\ea 
where $(n_s-1) \equiv d\ln \calP_{\calR_s}/d\ln k$ is the spectral index and $\tau_*$ is the horizon-crossing conformal time. We have used the conservation of the long mode to replace $\calR_\ell(\tilde \tau)$ with $\calR_\ell(0)$ and neglected the time dependence of $H$ to convert derivatives with respect to $a_*$ to derivatives with respect to $k$. The extra minus sign appears since in the presence of a positive long mode, the short mode with a fixed comoving wavenumber crosses the horizon slightly earlier \cite{Maldacena:2002vr}. Plugging this result into Eq.~\eqref{main} and noting that in slow-roll inflation $\calR_\ell$ is conserved in the entire range of integration we find
\ba 
\fNL = -\dfrac{5}{12}(n_s-1).
\ea 
 This is precisely the standard non-Gaussianity consistency condition in the slow-roll inflationary background. 

\subsection{Ultra-slow-roll inflation and the transition to a slow-roll phase}
\label{USR}
We now examine how the consistency relation Eq.~\eqref{main} works in the case of an ultra-slow-roll (USR) phase of inflation. The USR phase happens when the inflaton field rolls on a flat potential.  Since the USR phase leads to a rapid growth of the curvature perturbation, we assume it ends before the non-perturbative effects dominate and consider a sudden transition to a slow-roll phase where the curvature perturbation soon relaxes to a final value and freezes afterward. Following \cite{Cai:2018dkf}, we assume the field moves from large to small values and model the potential by 
\ba 
V(\phi) =  V_0 \left[ 1 + \sqrt{2 \epsilon_V} (\phi-\phi_e) \theta(\phi_e - \phi) /\Mpl \right]
\ea 
where the constant $V_0$ is the dominant term, $\phi_e$ is the field value at the transition and we assumed a slow-roll potential after $\phi_e$ with a constant slope that is controlled  by $\epsilon_v$ (which is the attractor slow-roll parameter).
Restricting the analysis to the modes that leave the horizon in the USR phase, we would need the mode function at  and after the transition which is given by \cite{Cai:2018dkf}
\ba 
\label{mode_func_USR}
\calR_k(\tau) \simeq  ( \frac{6-h}{6}+\frac{h\tau^3}{\tau_e^3}  ) \dfrac{H}{2\Mpl \sqrt{  \epsilon(\tau) k^3}} \, , \qquad \text{for $\tau_e\leq \tau$ and $k|\tau_e| \ll 1$}\, ,
\ea 
where $h\equiv -6\sqrt{\epsilon_v/\epsilon_e}$, the time variation of $H$ is neglected and the subscript $e$ denotes that the parameters are evaluated at the end of the USR phase, i.e., at the transition time.  Note that during the USR phase, the slow-roll parameter is given by $\epsilon=\frac{\dot \phi_i^2}{2 H^2 \Mplsq} (a/a_i)^{-6}$, where $\dot \phi_i$ is some initial velocity. After the transition $\epsilon$ relaxes to the attractor value $\epsilon_v$. The relaxation time is controlled by the value of $h$; for $h \to -\infty$ the relaxation is instant and the mode function freezes immediately after the transition. We will keep $h$ arbitrary in our analysis.  From Eq.~\eqref{mode_func_USR} it is evident that the late-time power spectrum is given by 
\ba 
\calP_{\calR_s} = \left(\dfrac{6-h}{6}\right)^2 \dfrac{H^2}{8\pi^2 \Mplsq  \epsilon_v}.
\ea 
This result depends on the history of the evolution through the initial conditions and through the dependence on the transition time $\tau_e$ (recall that $h$ depends on $\epsilon_e$ which is the slow-roll parameter at $\tau_e$) . The effect of the long mode on the initial conditions is negligible for the following reason: The power spectrum only depends on the initial velocity (and not on the initial field value) via $\epsilon$ which is given in the text below Eq.~\eqref{mode_func_USR}. A long mode induces a change to $\dot \phi_i$ via the time derivative of the relation $\Delta \phi =-(\dot \phi /H)\calR_\ell $.\footnote{This is the standard relation between inflaton's fluctuation and the curvature perturbation. In our convention, $\dot \phi$ is always negative. Therefore, this relation implies that a positive $\calR_\ell$ leads to a positive inflaton's fluctuation.} However, it is easy to check that this field fluctuation is almost conserved even in the USR phase,  $\partial \Delta \phi/\partial t$ decays rapidly and, therefore, cannot significantly alter $\dot \phi_i$  \cite{Namjoo:2012aa}. We are thus left with the effect of the long mode on the transition time. Again, this effect can be captured by noting that $\calR_\ell$ rescales the scale factor as seen by the short modes, i.e.,  $ a \to a\, e^{\calR_\ell}$. This results in
\ba 
\label{func_deriv_USR}
\calR_\ell(\tilde \tau)\dfrac{\delta }{\delta \calR_\ell(\tilde \tau) } \langle \calR_s^2(0) \rangle'  \simeq \calR_\ell(\tau_e) \dfrac{dP_{\calR_s}}{d\ln a_e}\delta(\tilde \tau-\tau_e).
\ea 
Note that, as opposed to the slow-roll case and the rescaling of the horizon-crossing time (Eq.~\eqref{func_deriv_SR}), we do not have a minus sign here. To see the difference, consider a positive long mode in an expanding universe that changes the scale factor from $a$ to $\hat  a =a\, e^{\calR_\ell}$ and a short mode with a fixed comoving wavenumber $k_s$. In the presence of the long mode, the short mode experiences the horizon-crossing time ($k_s=\hat a H$) slightly earlier. In contrast, the transition time happens slightly later since when the background inflaton field reaches $\phi_e$, the scale factor (hence the universe) would be slightly bigger.  Another crucial point to notice  in Eq.~\eqref{func_deriv_USR} is that $\calR_\ell(\tau_e)$ appears which is different from the final value $\calR_\ell(0)$ due to the large-scale evolution. According to Eq.~\eqref{mode_func_USR},  they are related via $\calR_\ell(0)=\frac{h-6}{h}\calR_\ell(\tau_e)$. This extra factor needs to be taken into account to obtain the correct results. Plugging Eq.~\eqref{func_deriv_USR} into the key relation for the bispectrum (Eq.~\eqref{main}) results in
\ba 
\label{fNL_USR}
\dfrac{12}{5}\fNL = \dfrac{h}{h-6} \dfrac{d\ln \calP_{\calR_s}}{d\ln a_e} = \dfrac{6 h^2}{(6-h)^2}.
\ea 
This is precisely the $\fNL$ obtained in \cite{Cai:2018dkf} by direct (in-in) computation. 
In taking derivatives with respect to $a_e$ one has to recall that $h$ is  a function of $a_e$ through $\epsilon_e$. In the instant relaxation limit $h \to -\infty$, one obtains $\fNL=5/2$ which is the size of non-Gaussianity before the transition. This simplified result is also obtained by other consistency relations based on the symmetries \cite{Bravo:2017wyw,Finelli:2017fml} which are broken in the presence of the transition (that is why the full result with a finite $h$ was not reproducible). To our knowledge, this is the first time that Eq.~\eqref{fNL_USR} is obtained by a consistency relation.\footnote{Ref.~\cite{Firouzjahi:2023xke} obtains the same result but claims that this holds where $\calP_{\calR_s}$ peaks. We obtain a different result for the peaks using our consistency relation (as well as by direct computations) which will be presented elsewhere \cite{peaks}.} 

\subsection{The non-attractor, non-canonical model   }

An even more non-trivial example is presented in \cite{Chen:2013aj} where a non-canonical inflaton field with a non-attractor evolution is considered. The Lagrangian is given by 
\ba 
L(X,\phi) = \dfrac{X^{\alpha}}{M^{4\alpha-4}} - V(\phi) \, , \qquad V(\phi)=V_0 + v \left(\dfrac{\phi }{\Mpl}\right)^{2\alpha}\, ,
\ea 
where $M$, $V_0$, $v$ and $\alpha$ are  constant parameters.
In this example, a peculiar background solution of the form $\phi \sim a^{\kappa}$ is found in \cite{Chen:2013aj} where $\kappa=-3/\alpha$ (the existence of that solution assumes that $V_0$ is the dominant term and determines $v$ in terms of other parameters). Similar to the USR model, the super-horizon curvature perturbation grows like $\calR_k \sim \epsilon^{-1/2} \sim a^{3}$. To simplify the analysis, we follow Ref.~\cite{Chen:2013aj} and assume an instant relaxation after a period of non-attractor behavior. Under this assumption, the final power spectrum is given by 
\ba 
\calP_{\calR_s} =\dfrac{H^2}{8\pi^2 \Mplsq \epsilon_e c_s}
\ea  
where $c_s^2=1/(2\alpha-1)$, $H$ is assumed to be a constant and $\epsilon_e$ is the slow-roll parameter computed at the end of the non-attractor phase. 

We take similar steps as in Secs.~\ref{SR} and \ref{USR} to obtain the squeezed limit bispectrum. In this example,  similar to the USR model, the effect of the long mode at the transition time must  be taken into account.  However, in contrast to our previous examples, the impact of the long mode on the initial condition also contributes to the final answer. 
 The reason for this difference is that the specified background solution is unstable under small perturbations. To take this effect into account, notice that the presence of  $\calR_{\ell}$ induces an inflaton fluctuation of the form $\Delta \phi = -\dot \phi \calR_{\ell}/H \sim a^{3+\kappa}$. Thus, a long-wavelength curvature perturbation at some early-time $\tau_i$ adds a correction to the background field as follows\footnote{This perturbative correction to the background field can also be obtained by perturbatively solving the background equations \cite{Chen:2013eea}.}
 \ba 
 \label{phi_correction}
 \phi \simeq \phi_i \, a^{\kappa} + \Delta \phi_i \,  a ^{3+\kappa} \simeq \phi_i a^{\kappa} \left( 1  - \kappa  \calR_{\ell ,i} a^{3}\right) \, ,
 \ea 
where $\phi_i$ is the unperturbed initial condition, $\calR_{\ell,i}$ is the long mode at $\tau_i$, we have chosen $a(\tau_i)=1$ and  the last equality is simplified using the unperturbed solution, since the linear perturbation suffices.
Crucially, note that the second term grows relative to the first term, indicating the instability of the background.  Using Eq.~\eqref{phi_correction}, up to linear order in $\calR_{\ell ,i}$, the slow-roll parameter receives a correction in the following way
\ba 
\label{epsilon_cs}
\epsilon \to  \epsilon \left (1+6(1-\alpha) \calR_{\ell,i} a^{3}\right).
\ea  
To summarize,  $\calR_\ell$ affects the power spectrum for the short modes through $\epsilon_e$ and it does so in two ways (and at two different instances of time): (i) it induces a  perturbation to the initial condition (Eq.~\eqref{epsilon_cs}) and (ii) it changes the transition time $\tau_e$ by rescaling the scale factor. Taking both effects into account results in
\ba 
\label{func_deriv_cs}
\calR_\ell(\tilde \tau)\dfrac{\delta }{\delta \calR_\ell(\tilde \tau) } \langle \calR_s^2(0) \rangle'  \simeq  \calR_\ell(\tau_i) \dfrac{dP_{\calR_s}}{d\calR_{\ell,i}}\delta(\tilde \tau-\tau_i) +\calR_\ell(\tau_e) \dfrac{dP_{\calR_s}}{d\ln a_e}\delta(\tilde \tau-\tau_e)\, ,
\ea 
Note that $\tau_i$ must be chosen to be in the range  $\tau_\ell < \tau_i \ll \tau_e $. I.e., we assume that  $\tau_i$ is early enough so that an even earlier effect of $\calR_\ell$ on $\calR_s$ (when $\calR_s$ is deep inside the horizon and resides in its vacuum) can be neglected. 
Using this result to simplify Eq.~\eqref{main} yields
\ba 
\label{fNL_cs}
\dfrac{12}{5} \fNL = -6(1-\alpha) + 6 = \dfrac{3(1+c_s^2)}{c_s^2}.
\ea 
The first term represents the impact of the long mode on the initial conditions, while the second term reflects its effect on the transition time. If we set $\alpha=c_s=1$, the former effect disappears and one recovers the USR result Eq.~\eqref{fNL_USR} in the instant relaxation ($h \to -\infty$) limit.  Note that the factor $a^3$  in the correction to $\epsilon$ from the perturbed initial conditions (Eq.~\eqref{epsilon_cs}) compensates the factor that appears in converting $\calR_{\ell }(\tau_i)$ (in Eq.~\eqref{main}) to $\calR_\ell(0)$ (that is needed for the usage of Eq.~\eqref{eq: fnl}).  Remarkably,  Eq.~\eqref{fNL_cs} is precisely the results obtained in \cite{Chen:2013eea} by direct computation of the three-point function. Again, to our knowledge, this is for the first time that this result is obtained using a consistency condition and the perturbation theory up to  linear order.\footnote{Ref.~\cite{Mooij:2015yka} obtains a consistency relation by symmetry considerations  in the case of non-canonical  inflation. However, to obtain the squeezed limit bispectrum, it is needed to go beyond the linear perturbations theory. Furthermore, despite the claims made in that paper,  their final result does not resemble  Eq.~\eqref{fNL_cs}. }

\section{Conclusion}
In this paper, we proposed a consistency relation that holds for all single-field models of inflation (Eq.~\eqref{main}). The reason for the generality is that the consistency relation does not rely on any assumption about the form of the Lagrangian governing inflation (other than that it is a single-field scenario) or about the specific solutions that are being considered. We reproduced the previously known results, some of which are obtained for the first time using a consistency relation and the perturbation theory up to  linear order.

There are several next steps worth considering.  Since our consistency relation is very generic, its violation is presumably not easy. Nonetheless, if such violations do exist, it would clarify the limitations and hidden assumptions behind our results that we may have overlooked.\footnote{We have experienced such a story for the consistency relation that holds for slow-roll inflation \cite{Namjoo:2012aa}.} Multiple-field models of inflation are probably the clearest examples where the violation is expected. It would also be  interesting to study the validity of this consistency condition in the presence of  spatial curvature which leads to the violation of some other consistency relations \cite{Avis:2019eav}. On the other hand,  extensions of the same ideas, e.g., to multiple-field setup or to higher order correlation functions are worth exploring.  We leave these further investigations to future works. 

We end with some remarks on the comparison between our method of computing the squeezed limit bispectrum and the $\delta N$ formalism. While both methods rely on the gradient expansion, we note that in our method  the gradient expansion is only needed for the long-wavelength mode. This is a milder assumption than what is needed for the $\delta N$ formalism where the validity of the gradient expansion is required to hold for all modes. Perhaps that is why we were able to reproduce the standard consistency relation in slow-roll inflation whereas the $\delta N$ formalism fails.\footnote{It is argued in \cite{Abolhasani:2018gyz} that a non-trivial amendment to the standard $\delta N$ formalism enables it to reproduce the correct consistency relation in slow-roll inflation.} On the other hand, our results for the USR and non-attractor inflation may be reproduced by the $\delta N$ formalism. However, one needs to go to the second order in perturbing the background solution while we only needed the linear perturbations. Finally, and more importantly, our method has a crucial advantage in that it allows the short modes to be sub-horizon at epochs that have a significant impact on the final statistics, such as phase transitions. This enables one to obtain  the full squeezed limit bispectrum as a function of $k_s$, e.g., including the scales at which peaks and troughs appear as a result of the sharp transition. Clearly, this is not achievable by the $\delta N$ formalism. We will study this problem in some detail in \cite{peaks} where we explore the important consequences of the consistency relation to the statistics of fluctuations at the scales corresponding to the peaks of the power spectrum. 

\acknowledgments
We thank Hassan Firouzjahi, Mehrdad Mirbabayi,  Bahar Nikbakht and Enrico Pajer for the fruitful discussions. 


\bibliography{refs} 
\bibliographystyle{JHEP}
%
%
%
%
\end{document}